# Dynamics of the verge and foliot clock escapement

P. Hoyng,[1,a)] and A.C.T. Nieuwenhuizen[1,b)]

[1] SRON Netherlands Institute for Space Research, Sorbonnelaan 2,
3584 CA Utrecht, The Netherlands

22 January 2018

The verge and foliot escapement has received relatively little attention in horology, despite the fact that it has been used in clocks for ages. We analyse the operation of a verge and foliot escapement in stationary swing. It is driven by a torque $m = \pm \mu$, switching sign at fixed swing angles $\pm \varphi_0$, and $\mu$ is taken to be constant. Friction is assumed to exert a torque proportional to the angular speed. We determine the shape of the swing angle $\varphi(t)$, and compute the period and the swing amplitude of the foliot analytically, as a function of the model parameters. We find that the period of the foliot scales as $P \propto \mu^{-1}$ for weak driving, gradually changing into $P \propto \mu^{-1/3}$ for strong driving (large $\mu$).

## I.  INTRODUCTION

Early mechanical clocks were generally equipped with a verge and foliot escapement, see Fig. 1. This mechanism to control the rate of a clock appeared around A.D. 1300 and has been used since then for hundreds of years[1]. By moving the small weights on the foliot the rate of the clock can be tuned. To our modern eyes, the foliot is a rather bad timekeeper. Some 15 min/day ($\approx 10^{-2}$) is about what you get. But in those days not yet ridden by notions of speed and efficiency, that was adequate for most purposes.

The physical reason for the low accuracy is that the verge plus foliot is a highly dissipative system. In each swing *all* the energy has to be fed anew into the foliot and taken out again by the driving mechanism and friction. The trouble is that it is difficult to make this supply and removal of energy sufficiently reproducible, whence sizeable variations and drifts in subsequent oscillation periods accumulate. At this point it is easy to sigh 'if they had only mounted some kind of a spring on the foliot - that would have improved the timekeeping significantly'. Yes, but that idea was only introduced in 1675 by Huygens.

Horologists did try to improve the performance, and arguably the most accurate foliot-equipped clocks have been made by Jost Bürgi (1552-1632), clockmaker and astronomer employed by Wilhelm IV landgrave of Hesse-Kassel[3, 4, 5]. His extant clocks are a marvel to see, and the craftsmanship with which they have been constructed makes you think they come straight from a modern mechanical workshop. They attain an accuracy of better than 1 min/day ($\approx 10^{-3}$) according to von Bassermann-Jordan[6], and represent the state of the art in horology around A.D. 1600. In actual fact Bürgi used a double-foliot or cross-beat escapement[7], but for the present that is a detail.

The foliot remained the clockmaker's workhorse until well into the second half of the 17th

a) Electronic mail: p.hoyng@sron.nl; peter.hoyng@gmail.com
b) Electronic mail: a.c.t.nieuwenhuizen@sron.nl

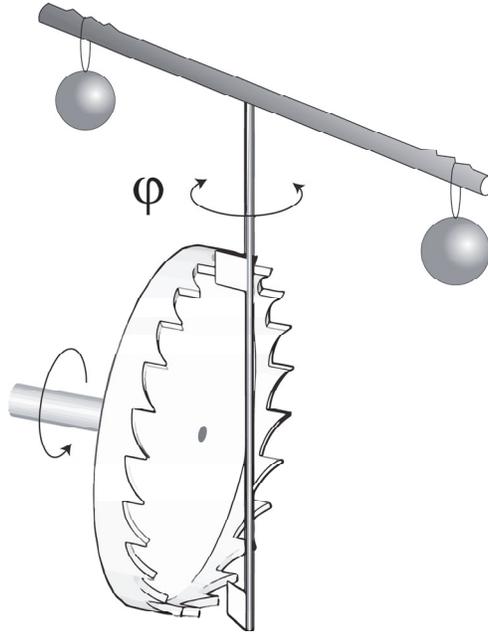

**FIG 1**: The verge and foliot escapement. The driving torque $m$ exerted on the verge switches sign after each half period, at fixed values $\pm\varphi_0$ of the swing angle $\varphi(t)$. (Adapted from Fig. 1 of Roup *et al.*[2])

century. After Huygens' 1657 invention it became obsolete, and was gradually replaced by a pendulum where and when the need arose. The great idea of a pendulum is that you don't supply and remove *all* the energy in each swing, but rather leave as much of it stored in the system as possible. In this way the pendulum is largely free in its motion (at least much more so than a foliot), and the escapement only supplies the fraction of the energy that is lost by friction. And only insofar energy is resupplied will inaccuracies and variability creep in. That is, in a nutshell, why pendulum clocks are able to keep time more accurately than verge-and-foliot clocks.

Theoretical investigations have been published by only a few authors. Lepschy et al.[8] analyse the verge and foliot as a two-body system whose parts are in continuous frictionless motion, interrupted by inelastic collisional impacts. Roup et al.[2] studied a comprehensive version of this model using impulsive differential equations. They determine under what conditions the system has a stable limit cycle and find that the period $P$ of the foliot scales as $\mu^{-1/2}$ where $\mu$ is the driving torque. Denny obtains the same scaling using a much simpler model[9]. It is not clear to what extent this is a coincidence or a more general result, as neither paper investigates the origin of their $\mu^{-1/2}$ scaling.

The shape of $\varphi(t)$ and its dependence on the parameters (driving torque $\mu$ and friction coefficient $a$) is an interesting topic in its own right that, to our knowledge, has not been studied before. We therefore consider the dynamics of a foliot driven by a constant torque, and friction proportional to angular velocity. We see this as a first step, and reserve other types of friction, such as Coulomb friction, for a later study. We introduce a model escapement and analyse the operation of this device in its historic context. In Sec. III we discuss the properties of the foliot and we review our results in Sec. IV.

## II. DYNAMICS OF A STATIONARY OSCILLATING FOLIOT

The foliot rotates on the vertically suspended verge and experiences torques due to driving and friction, see Fig. 1. The swing angle $\varphi$ obeys the same differential equation as that of a pendulum, except that there is no restoring gravity torque:

$$\ddot{\varphi} + a\dot{\varphi} = m(\varphi) . \tag{1}$$

Here $a$ is the friction coefficient (dimension $[a] = $ s$^{-1}$) and $m$ the driving torque, more precisely, $m$ = driving torque / (moment of inertia of the foliot). The dot stands for the time-derivative: $\dot{} = d/dt$, and $\ddot{} = d^2/dt^2$. The escapement delivers a driving torque $m = \mu$ that switches to the opposite sign $m = -\mu$ after half a period $p = P/2$ (we take $\mu > 0$). The switch is at fixed, mechanically determined angles $\pm \varphi_0$. For convenience we assume that $a$ and $\mu$ do not depend on $\varphi$ and $\dot{\varphi}$. It seems plausible that the dynamics of the foliot on time scales of a period and longer are not very sensitive to the fine structure of $m(\varphi)$, as in the case of a pendulum[10]. Hence, we adopt this simple $m = \pm\mu$ flip-flop model.

Eq. (1) can be further simplified by introducing a dimensionless time $\tau = at$, and a normalised swing angle $\vartheta = \varphi/\varphi_0$ leading to

$$\vartheta'' + \vartheta' = \pm x , \qquad x = \mu/(a^2 \varphi_0) , \tag{2}$$

where $' = d/d\tau$, etc. But the advantage is marginal and we mention (2) only because it nicely illustrates why the normalised torque $x$ figures so prominently below.

### A. ANALYSIS

We choose $t = 0$ at the moment when $m$ switches sign from $+\mu$ to $-\mu$, see Fig. 2, and take as initial conditions

$$\varphi(0) = \varphi_0 ; \qquad \dot{\varphi}(0) = \dot{\varphi}_0 . \tag{3}$$

Here $\varphi_0$ and $\dot{\varphi}_0$ are the initial position and angular velocity. Assuming that $m = -\mu$ and that $a$ and $\mu$ do not depend on time, we may solve the differential equation (1):

$$\varphi(t) = \varphi_0 + \frac{\mu + a\dot{\varphi}_0}{a^2}(1 - e^{-at}) - \frac{\mu t}{a} , \tag{4}$$

valid for $0 \leq t \leq p$, where $p$ is the moment of the next sign flip of $m$. The correctness of (4) may be readily ascertained by back-substitution in (1) and performing the differentations. For a stationary swinging foliot as we assume here, $p$ is also the half period $P/2$. Since $\varphi_0$, $a$ and $\mu$ are known model parameters, relation (4) fixes the motion of the foliot as soon as we know $\dot{\varphi}_0$. Its value may be found by noting that after half a period $p$, and in a stationary

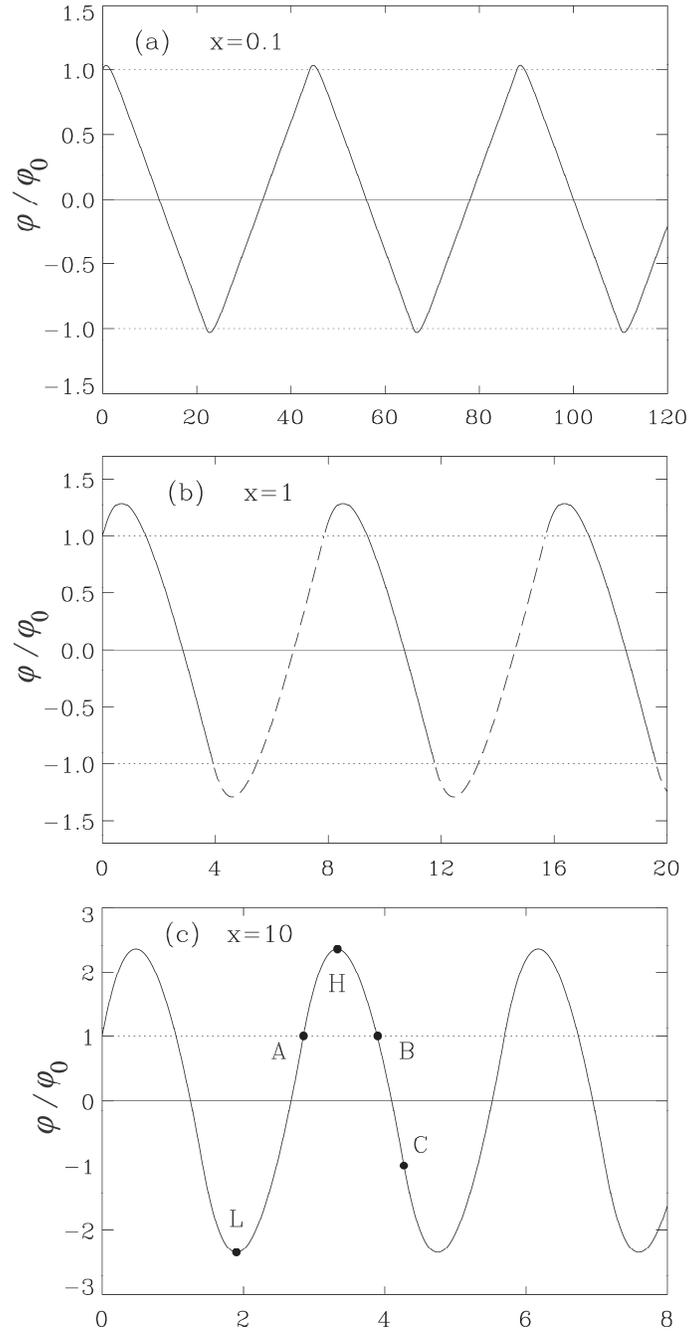

**FIG** 2: From top to bottom: sample $\varphi(t)$ for small, medium and large normalised driving torque $x = \mu/(a^2\varphi_0)$. On the horizontal axes is dimensionless time $at$. It is evident that the stronger the driving, the smaller the period and the larger the amplitude $\varphi_m$. The middle panel shows how $\varphi(t)$ can be constructed by pasting together pieces of expression (4). The points $L, A, H, B$ and $C$ in panel (c) are referred to in the text.

state, the swing angle $\varphi(p)$ and its derivative must assume values opposite to those in (3), so

$$\varphi(p) = -\varphi_0, \quad \dot{\varphi}(p) = -\dot{\varphi}_0. \tag{5}$$

With the help of expression (4) this may be written as:

$$\frac{\mu + a\dot{\varphi}_0}{a^2}(1 - e^{-ap}) - \frac{\mu p}{a} + 2\varphi_0 = 0, \tag{6}$$

$$(\mu + a\dot{\varphi}_0)e^{-ap} = \mu - a\dot{\varphi}_0. \tag{7}$$

These relations (6) and (7) are two equations that determine the values of $\dot{\varphi}_0$ and $p$, and we show in Appendix A how they may be computed.

Supposing that that has been done, we may then compute $\varphi(t)$ for all $t$ by gluing together pieces of expression (4) or its dimensionless form (A7) lasting half a period $p$ with alternating sign, as shown in Fig. 2b. For example, for $p \leq t \leq 2p$ we have $\varphi(t) = -\varphi(t - p)$, and so on. This construction guarantees that $\varphi$ is a smooth function of $t$. We may also compute the swing amplitude $\varphi_m$ and the period $P = 2p$, and other desired quantities. We relegate the technicalities to Appendix A, and restrict ourselves below to a discussion of the results.

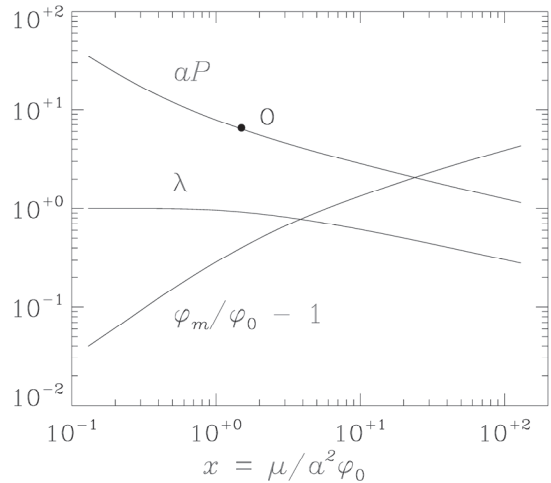

**FIG 3**: Period $aP$ and amplitude $\varphi_m/\varphi_0$ of the foliot, and the auxiliary parameter $\lambda = a\dot{\varphi}_0/\mu$ (= the characteristic dimensionless angular speed) as a function of the normalised driving torque $x = \mu/(a^2\varphi_0)$. These are obtained as follows. Given a value of $x$, we solve $\lambda$ from (A4) as outlined in Appendix A. Period and amplitude then follow from (A5) and (A6). Point $O$ is referred to in Sec. III.B.

### III. PROPERTIES OF THE STATIONARY MOTION

Fig. 2 shows the angle $\varphi(t)$ of a foliot in stationary swing for several driving torques. For weak driving ($x \ll 1$; top panel) the foliot moves slowly at virtually constant speed $\dot\varphi = \mu/a$, as inertia can be ignored in Eq. (1) and there is hardly any overshoot at the turning points. So the period is approximately equal to $P \approx 4\varphi_0/(\mu/a) = 4a\varphi_0/\mu$, or in terms of the normalised torque $x$: $aP \approx 4/x$, as displayed in Table 1. It may take ages to complete a period when $\mu$ is small, but there is no minimum torque. The foliot is, at least theoretically, always self-starting and the amplitude $\varphi_m$ is always larger than $\varphi_0$.

For moderate driving (middle panel), the swing angle overshoots $\varphi_0$, by about 30% for $x = 1$. The period has decreased significantly and $\varphi(t)$ has developed a noticeable asymmetry. This is due to the fact that when the swing angle $\varphi$ reaches position $A$ in Fig. 2c, the foliot has traversed an acceleration traject ($LA$) that is longer than $HB$ when it arrives in $B$. Consequently, the angular speed $\dot\varphi$ in $A$ is larger than in $B$, and that makes that the peaks appear to 'recline'.

For strong driving ($x \gg 1$; bottom panel) the period decreases further, and the overshoot is large. The peaks are also rounder than they would be for sinusoidal motion. Fig. 2 illustrates that the motion of the foliot is neither harmonic, nor isochronous as the period depends on $\mu$.

### A. PERIOD AND AMPLITUDE

The analysis in Appendix A shows that

1. the dimensionless period $aP$ and the amplitude $\varphi_m/\varphi_0$ depend solely on $x = \mu/(a^2\varphi_0)$;
2. values of $P$ and $\varphi_m/\varphi_0$ may be computed numerically as outlined in Appendix A and displayed in Fig. 3;
3. in the limit of weak and strong driving asymptotic expressions are available. These are derived in Appendix B, and collected in Table 1.

TABLE 1: Relative amplitude $\varphi_m/\varphi_0$ and period $aP$ for small and large driving torque $x = \mu/(a^2\varphi_0)$.

|           | $\lambda = a\dot\varphi_0/\mu$ | $\varphi_m/\varphi_0 - 1$ | $aP$ |
|-----------|------------------------------|---------------------------|------|
| $x \ll 1$ | $1$                          | $(1-\log 2)x$             | $4/x$ |
| $x \gg 1$ | $\left(\dfrac{x}{3}\right)^{-1/3}$ | $\dfrac{3}{2}\left(\dfrac{x}{3}\right)^{1/3}$ | $4\left(\dfrac{x}{3}\right)^{-1/3}$ |

For example, for strong driving ($x \gg 1$) we read from Table 1 that $aP \approx 4(x/3)^{-1/3}$. Restoring the physical dimensions with $x = \mu/(a^2\varphi_0)$ we get:

$$P \approx 4(3\varphi_0/\mu a)^{1/3} \approx 5.8(\varphi_0/\mu a)^{1/3}. \tag{8}$$

We may summarize our results for the scaling of the period $P$ with $\mu$ as follows:

$$P \propto \mu^{-\gamma}; \qquad \frac{1}{3} < \gamma < 1. \tag{9}$$

The period scales as $\mu^{-1}$ for small $\mu$, and changes gradually into $P \propto \mu^{-1/3}$ as the driving gets strong. On the other hand, Roup et al.[2] and Denny[9] find $\gamma = 0.5$, apparently for all $\mu$. The origin of this difference remains to be investigated.

## B. SENSITIVITY TO PERTURBATIONS

As an example of how Fig. 3 may be used we study the sensitivity of the period to variations in driving torque and friction. We make a local power law fit to the $aP$ curve in Fig. 3 by writing $P = const \cdot a^{-1} x^{-\gamma}$. Starting from $\delta P = (\partial P/\partial \mu)\delta \mu + (\partial P/\partial a)\delta a$ and using that $dx/d\mu = x/\mu$ and $dx/da = -2x/a$, plus a little algebra, we arrive at

$$\frac{\delta P}{P} = -\gamma \frac{\delta \mu}{\mu} + (2\gamma - 1) \frac{\delta a}{a}. \tag{10}$$

Hence, $\gamma = 0.5$ (or $P \propto \mu^{-1/2}$) seems to be a good operating point, as the period is then insensitive to variations in friction. The corresponding value of $x$ is obtained by constructing the tangent to the $aP$ curve in Fig. 3 with inclination $\gamma = 0.5$. In this way we arrive at point $O$ at $x \approx 1.5$, so that $\varphi_m/\varphi_0 \approx 1.4$ and $P \approx 6.4/a$ (these numbers were simply read from the figure). Unfortunately it is not possible to eliminate the sensitivity to driving torque variations $\delta \mu$.

We conclude from (10) that the $10^{-2}$ timing accuracy quoted in Sec. I requires a driving torque stability of $3\%$ when the driving is strong ($x \gg 1$) and $1\%$ for weak driving ($x \ll 1$). In this model, strong driving makes a verge-and-foliot clock a more accurate time keeper (less sensitive to driving torque variations) than weak driving.

## IV. DISCUSSION AND SUMMARY

The verge plus foliot escapement has received much less attention in the literature than the pendulum. This is unfortunate because the foliot poses an interesting problem in theoretical horology. We have studied the dynamics of a foliot performing a stationary swing, assuming a constant driving torque and friction proportional to angular speed. The merit of this model is that most of the analysis can be done analytically.

The analysis is straightforward and one may wonder why it has not been done earlier. In part, the reason must be that by the time analyses of the kind presented here became possible, say around A.D. 1700, the perfection of the foliot had long since been completed empirically by horologists such as Jost Bürgi and his peers. So there may never have been a need for it.

We constructed the swing angle $\varphi(t)$ by smoothly pasting together pieces lasting half a period. The resulting $\varphi(t)$ has a characteristic asymmetry in that the peaks are 'leaning backwards.' We developed a method to compute the period $P$ and the amplitude $\varphi_m/\varphi_0$ numerically for given parameters. We find that the period of the foliot scales as $P \propto \mu^{-1}$ for weak driving, slowly changing into $P \propto \mu^{-1/3}$ for large $\mu$. In this strong driving limit period and amplitude change surprisingly slowly with the driving torque. For example, a tenfold larger driving torque $\mu$ reduces the period by a factor $10^{-1/3} \approx 0.5$ and makes the amplitude a factor $10^{1/3} \approx 2$ larger.

The reason for assuming that $a$ and $\mu$ are constant is foremost our wish to keep things simple. We suspect that the time keeping properties of the foliot depend mainly on some average of $m$ and $a$ over a period, as in the case of a pendulum.[10] To prove this for a verge-and-foliot escapement requires averaging the equation of motion (1) over a period, which is not straightforward as the foliot has no well-determined period. So for now it is merely a plausible assumption.

One should certainly question the idea of $a = $ constant, in view of the sensitivity of the motion to variations in friction, in particular when $\mu$ and/or the angular speed $\dot\varphi$ are small. Friction may change its type and switch to Coulomb friction, for example. This in turn will affect the self-starting property. Problems of this nature are best tackled with the help of numerical simulations of equation (1) with a variable friction $a$ and/or driving torque $\mu$. That will help to develop a more complete picture of the behaviour of the foliot under various circumstances.

**ACKNOWLEDGEMENTS**

**APPENDIX A: COMPUTATION OF THE PERIOD AND AMPLITUDE**

We begin by solving $e^{-ap}$ from (7), and substitute that in (6). After some algebra we find

$$p = \frac{2}{\mu}(\dot\varphi_0 + a\varphi_0). \qquad (A1)$$

The next step is to write (7) as

$$ap = \log\left(\frac{\mu + a\dot\varphi_0}{\mu - a\dot\varphi_0}\right), \qquad (A2)$$

and to eliminate $p$ on the left with relation (A1):

$$\log\left(\frac{\mu + a\dot{\varphi}_0}{\mu - a\dot{\varphi}_0}\right) - \frac{2a\dot{\varphi}_0}{\mu} = \frac{2a^2\varphi_0}{\mu}. \tag{A3}$$

This relation determines $\dot{\varphi}_0$ for given $a$, $\mu$ and $\varphi_0$. We reformulate it as follows. The parameter $\lambda = a\dot{\varphi}_0/\mu$ must be solved from the equation

$$f(\lambda) = \frac{2}{x}, \quad \text{where} \quad f(\lambda) = \log\left(\frac{1+\lambda}{1-\lambda}\right) - 2\lambda \approx \frac{2\lambda^3}{3}. \tag{A4}$$

The last expression, $f(\lambda) \approx 2\lambda^3/3$, holds for $\lambda \ll 1$. It can be shown that for positive $x$ there is always one root $\lambda$ in the interval $0 < \lambda < 1$. The easiest way to find it is by repeated division of this interval.

Then we need the expressions for the period and the amplitude. The time when $\varphi(t)$ attains a maximum is found by setting the time derivative of expression (4) to zero, after which the amplitude $\varphi_m$ follows by back substitution in (4):

$$\frac{\varphi_m}{\varphi_0} - 1 = x[\lambda - \log(1+\lambda)]. \tag{A5}$$

For the period $P = 2p$ we obtain with relation (A1):

$$P = \frac{4}{a}(\lambda + x^{-1}). \tag{A6}$$

Finally we mention the dimensionless form of expression (4):

$$\frac{\varphi(t)}{\varphi_0} = 1 + x\left[(1+\lambda)(1-e^{-at}) - at\right]. \tag{A7}$$

**APPENDIX B: ASYMPTOTIC SCALING**

We compute the asymptotic scaling of the auxiliary parameter $\lambda$, the period $P$ and the amplitude $\varphi_m$ as a function of $x$. When $x \ll 1$ we infer from (A4) that $\lambda \approx 1$, and we may simply set $\lambda = 1$ in (A5) and (A6). The limit $x \gg 1$ needs more work. In that case (A4) tells us that $\lambda$ is small, so we may expand $f(\lambda)$ to find that $2\lambda^3/3 \approx 2/x$, i.e. $\lambda \approx (x/3)^{-1/3}$. Here we see, incidentally, how the fractional exponents $\pm 1/3$ emerge in the theory. Next we expand in (A5) $\log(1+\lambda) \approx \lambda - \lambda^2/2$. Result:

$$\frac{\varphi_m}{\varphi_0} - 1 \approx \frac{1}{2}x\lambda^2 \approx \frac{1}{2}x\left[(x/3)^{-1/3}\right]^2 = (3/2)(x/3)^{1/3}. \tag{B1}$$

And for the period we get: $aP \approx 4\left[(x/3)^{-1/3} + x^{-1}\right] \approx 4(x/3)^{-1/3}$. These scalings have been summarised in Table 1.